# Cold-burst method for nanoparticle formation with natural triglyceride oils

**Diana Cholakova, Desislava Glushkova, Slavka Tcholakova, Nikolai Denkov***

*Department of Chemical and Pharmaceutical Engineering*
*Faculty of Chemistry and Pharmacy, Sofia University,*
*1 James Bourchier Avenue, 1164 Sofia, Bulgaria*

*Corresponding authors:
Prof. Nikolai Denkov
Department of Chemical and Pharmaceutical Engineering
Sofia University
1 James Bourchier Ave.,
Sofia 1164
Bulgaria
E-mail: nd@lcpe.uni-sofia.bg
Tel: +359 2 8161639
Fax: +359 2 9625643

**ABSTRACT**

Preparation of nanoemulsions of triglyceride oils in water usually requires high mechanical energy and sophisticated equipment. Recently, we showed that α-to-β (viz. gel-to-crystal) phase transition, observed with most lipid substances (triglycerides, diglycerides, phospholipids, alkanes, etc.), may cause spontaneous disintegration of micro-particles of these lipids, dispersed in aqueous solutions of appropriate surfactants, into nanometer particles/drops using a simple cooling/heating cycle of the lipid dispersion (Cholakova et al. *ACS Nano* **14** (2020) 8594). In the current study we show that this "cold-burst process" is observed also with natural oils of high practical interest, incl. coconut oil, palm kernel oil and cocoa butter. Mean drop diameters of ca. 50 to 100 nm were achieved with some of the studied oils. From the results of dedicated model experiments we conclude that intensive nano-fragmentation is observed when the following requirements are met: (1) The three phase contact angle at the air-water-solid lipid interface is below *ca.* 30 degrees; (2) The equilibrium surface tension of the surfactant solution is below *ca.* 30 mN/m and the dynamic surface tension decreases rapidly. (3) The surfactant solution contains non-spherical surfactant micelles. *e.g.* ellipsoidal micelles or bigger supramolecular aggregates; (4) The three phase contact angle measured at the contact line (frozen oil-melted oil-surfactant solution) is also relatively low. The mechanism(s) of the particle bursting process is revealed and, on this basis, the role of all these factors is clarified and discussed. We explain all main effects observed experimentally and define guiding principles for optimization of the cold-burst process in various, practically relevant lipid-surfactant systems.

**INTRODUCTION**

The production of submicrometer emulsion droplets usually requires high mechanical energy. Microfluidisers and high pressure homogenizers are used in which most of the energy is lost as heat and sound - less than 0.01% of the energy input is consumed for the actual drop break up and the related increase of drop surface area and energy.[1] Several low-energy methods were proposed in the literature, incl. phase inversion temperature, phase inversion composition, solvent diffusion method, and self-emulsification upon cooling and/or heating of alkane-in-water emulsions, each of them having some advantages and disadvantages.[2-6]

Recently, we discovered a new efficient method for spontaneous bursting of lipid particles (so-called "cold-burst" process) which resulted in a drop size decrease from ca. 100 μm to 0.4 μm after one cooling and heating cycle only of the lipid dispersion, see Figure 1 and Movie 1 for illustrative examples.[7] Briefly, the mechanism of the cold-burst process is the following. Upon fast cooling, many lipids, e.g. triacylglycerols (triglycerides, viz. esters derived from glycerol and three fatty acids) or diacylglycerols (diglycerides, esters derived from glycerol and two fatty acids), crystallize into thermodynamically meta-stable α-polymorph phase which has hexagonal crystalline lattice. Upon storage at low temperature (solid-state phase transition) or upon heating (melt-mediated phase transition) the crystalline order changes to thermodynamically more stable β-phase, Figure 2a.[8-11] This phase transformation is accompanied by local shrinkage of the lipid phase, thus leading to formation of "nanovoids" between the newly formed crystalline domains of the β-phase. In other words, after the α→β phase transition, the lipid phase acquires internal structure of crystalline domains, separated by nanoporous network.[7] The formed nanovoids were reported to have reduced pressure (so-called "negative pressure effect").[12-14] In aqueous surfactant solutions, this negative pressure sucks the surrounding aqueous phase.[7] Depending on the specific surfactants, upon storage or upon heating, the lipid particles increase their volume and, finally, each of them disintegrates (bursts) into millions of smaller particulates.[7]

In our previous study we found that the cold-burst process is observed with surfactants ensuring low three-phase contact angle at the solid triglyceride-water-air contact line, *ca.* < 50-60°, whereas for systems with high contact angle (> 90-100°), we observed entrapment of the penetrated water inside the lipid globule in the moment of its melting, with resulting formation of double emulsion containing water-in-oil-in-water drops.[7]

In our first study, the cold-burst process was demonstrated with pure triglycerides (TG), several binary and ternary TG mixtures, diglycerides and alkanes.[7] However, the natural triglycerides with high content of saturated alkyl chains, such as coconut oil (CNO), palm kernel oil (PKO), cocoa butter (CB) and lard are very complex mixtures of triglycerides which contain many different alkyl chains. For example, the typical chain-length distribution in the coconut oil is: 48.2 % $C_{12}$, 19.5 % $C_{14}$, 9 % $C_{16}$, 7.9 % $C_8$, 6 % $C_{18:1}$, 5.8 % $C_{10}$ and 3.6 % others.[15-17] About 22 % of the triglycerides in CNO consist of trilaurin with three $C_{12}$ chains, *ca.* 33 % contain two lauric chains ($C_{12}$) and one different chain ($C_{10}$ or $C_{14}$), *ca.* 15 % have two capric chains ($C_{10}$) and one lauric chain, 10 % have two myristic chains ($C_{14}$) and one lauric chain, and the remaining 20 % contain various other combinations of chain, see Supplementary Table S1 for detailed composition of the studied oils.[15-19] Therefore, it is not obvious in advance whether the freezing pattern of such complex triglyceride mixtures, containing components with very wide range of melting temperatures, would create nanoporous structure able to sustain negative pressure and to undergo cold-burst fragmentation.

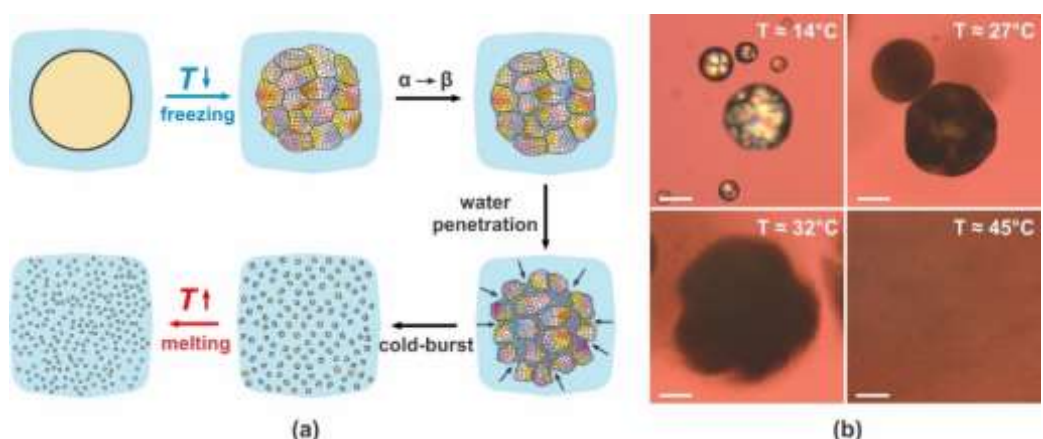

**Figure 1. Cold-burst process observed with pure triglycerides and model triglyceride mixtures.** (a) Schematics of the mechanism of cold-bursting process. Upon cooling, TG drops freeze in meta-stable α-polymorph which then transforms into the stable β-polymorph. This transition is accompanied by volume shrinkage of the crystalline domains and formation of nanoporous structure inside the lipid particle. Upon prolonged storage or upon heating, the surrounding aqueous phase penetrates into the frozen particles and fragments them into millions of nanometer-sized lipid particles. Upon further heating, these particles melt forming nanodroplets. This scheme is adapted from Fig. 3 in Ref. [7]. (b) Illustration of the cold-burst as observed with trilaurin drops ($C_{12}TG$) dispersed in aqueous solution of 1.5 wt. % $C_{12}SorbEO_{20}$ and 0.5 wt. % $C_{18:1}EO_2$ surfactants. Upon heating, the aqueous phase penetrates inside the particles which increase their volume. At 27°C the small lipid particles have already disintegrated and, upon further heating up to 45°C, the bigger lipid particles disintegrate completely as well. Scale bars, 50 μm.

From the literature we know that, upon cooling, the CNO first freezes in α-phase with lamella spacing of 3.84 nm and WAXD peak at 0.412 nm.[20] The β' polymorph with lamella distance 3.29 nm and short spacing at 0.429 nm and 0.383 nm appears at lower temperatures and co-exists with the α-polymorph.[20] These phases are observed by DSC analysis to crystallize separately upon quick cooling of CNO, see Figure 2b. Upon heating, the two polymorphs transform into another β' phase with lamellar spacing of 3.35 nm and three WAXD peaks at 0.430, 0.416 and 0.384 nm.[20] This polymorph transformation is also observed in the DSC thermograms as a shoulder peak, which is followed by a broad peak reflecting the complete melting of the coconut oil, see Figure 2.

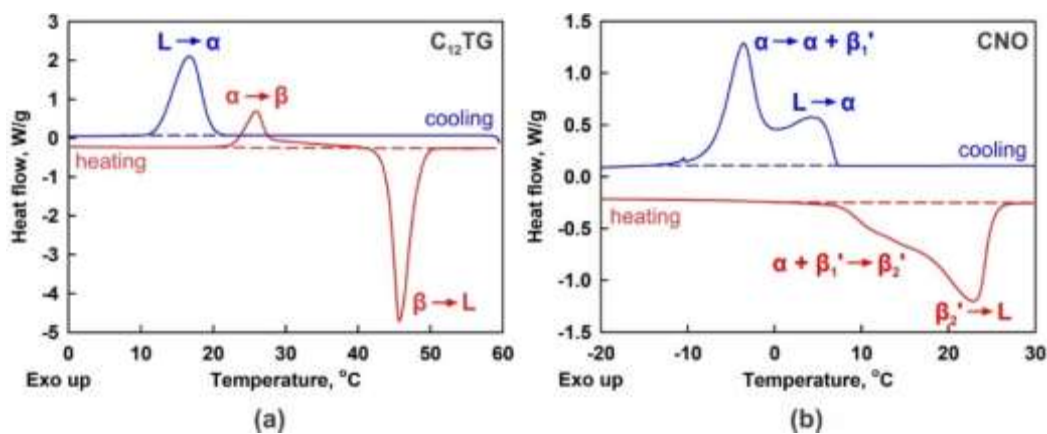

**Figure 2. DSC thermograms for bulk $C_{12}TG$ (a) and CNO (b).** Notations on the graphs show the phase transitions corresponding to the observed peaks, as explained in the text. The blue curves show the peaks observed upon 5°C/min cooling and the red curves – the peaks upon 5°C/min heating. With "L" we denote the isotropic liquid phase. First, the cooling curves are obtained and, afterwards, the same samples are heated until their complete melting.

The appearance of polymorphs in CNO opens the possibility for observing the cold-burst process with natural triglyceride oils, but the conditions for its realization are unclear. In the current study we explore this possibility and define the guiding rules for optimization of this process. We show that both the total surfactant concentration and the ratio of oil-soluble to water-soluble surfactant in the aqueous phase play key roles for the process efficiency. Furthermore, some minor components present in the technical grade surfactants were found to play a role. We reveal the mechanisms, explain the observed trends and show that the main findings are general and can be applied to other natural oils (cocoa butter, palm kernel oil, *etc.*).

We have chosen CNO for the systematic part of the study, because it is well characterized in literature and widely used in various industrial applications: for example in foods for preparation of chocolate, ice cream and other confectionary products;[21-24] in cosmetics as a key ingredient in creams, shampoos, lip glosses, body moisturizers, *etc.*;[25,26] in fuel industry as an alternative source for production of biodiesel and as environmental friendly lubricant;[27,28] and in pharmacy as a solubilizing excipient for oral and injectable formulations.[29,30] Various studies suggested also that CNO may provide health benefits, such as hearth health support, memory enhancement, improvement in immunity, and possibly prevention/delay of the development of Alzheimer's and Parkinson's diseases.[31,32]

**EXPERIMENTAL SECTION**

**Materials**

*Oils*

For preparation of the emulsions we used coconut oil (CNO) produced by Smart Organic, Bulgaria (under the brand Dragon Superfoods), purchased from local grocery store. We tested also several other sources of coconut oil, including coconut oil from *Cocos nucifera*, purchased from Sigma-Aldrich. We did not observe any significant differences between the samples prepared from different CNO sources. We studied also palm kernel oil (denoted as PKO, produced by BiOrigins)[18-19] and cocoa butter (CB, Dragon Superfoods)[33,34]. CNO, PKO and CB are naturally occurring triglyceride compounds with mixed triglyceride chains and predominant content of saturated alkyl chains, see Supplementary Table S1.

In another series of experiments we studied commercially available Precirol ATO 5 (PRE) and Gelucire 43/01 (GEL01) oils, both products of Gattefosse. These compounds are usually applied in pharmaceutical products for preparation of solid lipid nanoparticles, nanocarriers, matrices for sustained release, and coatings for protection and taste masking.[35-38] PRE is a mixture of $C_{16}$-$C_{18}$ monoglycerides (MG), diglycerides (DG) and triglycerides (TG), with predominant DG fraction (MG:DG:TG = 21:54:25 as we determined by gas chromatography). GEL01 contains triglycerides only and has similar composition to CNO, but with higher content of longer alkyl chains, see Supplementary Table S1.

In control experiments, we used glyceryl tridodecanoate (trilaurin) oil and tritetradecanoate (trimyristin) with purity > 98 % and > 95 % respectively, purchased from TCI Chemicals. In the three-phase contact angle measurements we used also medium chain triglyceride oil (Kollisolv® MCT 70), purchased from BASF.

All studied oils are approved for food, cosmetic and pharmaceutical applications.

*Surfactants*

For emulsion stabilization we used various water-soluble and oil-soluble surfactants. A detailed description of their chemical structures, producers and properties is presented in Supplementary Table S2. The tested nonionic water-soluble surfactants include polyoxyethylene alkyl ethers with general formula $C_nEO_m$ (from the series with trade name Brij or Lutensol), with hydrophobic chain length, $n$, varied between 8 and 18 C-atoms and number of ethoxy units, $m$, varied between 2 and 23. We tested also different polyoxyethylene sorbitan monoalkylates, $C_nSorbEO_{20}$ (trade name Tween). We used also four anionic surfactants: two sodium alkyl sulfates, $CH_3(CH_2)_7SO_4Na$ and $CH_3(CH_2)_{11}SO_4Na$, SDS, sodium lauryl ether sulfate with one ethoxy group (SLES), sodium dodecylbenzenesulfonate (LAS), and one amphoteric surfactant – cocoamidopropylbetaine (CAPB).

Several series of oil-soluble surfactants and cosurfactants were also studied. They include three surfactants from the Brij series, $C_nEO_m$, with *n* varied between 12 and 18, and *m* varied between 2 and 4. Polyoxyethyleneglycol monooleylether with 2 EO units, $C_{18:1}EO_2$ (TCI Chemicals), and sorbitan monoalkylates from Span series, denoted as $C_nSorb$ in the text, were also tested. We used monoacylglycerides with different chain lengths: $C_8MG$ to $C_{18}MG$ and the unsaturated monooleinglyceride, $C_{18:1}MG$. In several experiments we used fatty acids or fatty alcohols ($C_nAc$, $C_nOH$) as cosurfactants.

All substances were used as received, except for 1-oleoyl-rac-glycerol > 40 %, purchased by Sigma-Aldrich, which was purified as described in the Methods section below. The aqueous solutions were prepared with deionized water with resistivity > 18 MΩ·cm, purified by Elix 3 module (Millipore).

Other chemicals

The anhydrous calcium chloride (purity 99.5%) was a product of Ferak Berlin GmbH. The hydrochloric acid, potassium hydroxide, ethanol and ethylene glycol were purchased from Teokom (Bulgaria) and were of analytical grade.

**Methods**

*Emulsions preparation*

The initial polydisperse oil-in-water emulsions, containing micrometer drops, were prepared via rotor-stator homogenization with Ultra Turrax (IKA, Germany). In some experiments, we prepared monodisperse emulsion by membrane emulsification, passing the oily phase through porous glass membranes.[39,40] We used a laboratory Microkit membrane emulsification module from Shirasu Porous Glass Technology (SPG, Miyazaki, Japan), working with tubular glass membranes of outer diameter 10 mm and working area of approx. 3 cm$^2$. The temperature was kept sufficiently high to maintain the liquid state of all substances during the emulsification process.

*Optical observations in a glass capillary*

Observations by optical microscopy were performed with specimens of the studied emulsions, placed in glass capillaries with length of 50 mm and rectangular cross-section: width of 1 mm or 2 mm and height of 0.1 mm. These capillaries were enclosed inside a custom-made thermostating aluminum chamber, with several orifices cut-out for optical observation. The chamber temperature was controlled by cryo-thermostat (JULABO CF30). The temperature in the chamber was measured with a thermo-couple probe, with an accuracy of ± 0.2°C, and calibrated with a precise mercury thermometer. The thermo-probe was inserted in one of the orifices of the thermostating chamber, where a capillary with emulsion sample would be normally positioned. The actual capillaries with specimens of the studied emulsions were placed in the neighboring orifices for optical observations. In control experiments, dispersions containing lipid micro-particles were heated to observe their melting. The melting process was always observed at temperatures very close, within ± 0.2°C, to the reported melting temperature of the bulk oil, $T_m$.

These observations were performed on an AxioImager.M2m microscope (Zeiss, Germany) in transmitted, cross-polarized white light, with included λ-compensator plate situated after the sample and before the analyzer, at 45° with respect to both the analyzer and the polarizer. The images used for determination of the initial drop size were made in transmitted white light. Long-focus objectives ×10, ×20, ×50 and ×100 were used.

*Three phase contact angle measurements*

The solid lipid substrates, used for measurements of the contact angles, were prepared by placing a small amount of melted $C_{14}TG$, CNO, or CNO+0.5 wt.% $C_{18:1}MG$ onto a pre-hydrophobized glass slide (the pre-hydrophobization was made with hexamethyldisilazane, HMDS). After that a second glass slide was placed on top of the melted oil to form a lipid layer of homogeneous thickness. Afterwards, these lipid layers were crystallized at 0°C for CNO and at 20°C for $C_{14}TG$, and stored in a freezer at -18°C prior to the actual contact angle experiments.

The three-phase contact angles at the water-air-solid lipid interface were measured as a function of the temperature with DSA 30 and DSA 100E apparatus (Krüss, Germany). These experiments were made by placing a small drop of the studied surfactant solution on the top surface of the CNO substrate. The temperature was controlled by TC40 thermostatic cell and measured with calibrated thermocouple. The experiments were performed from an initial temperature of $T \approx 7°C$ up to ca. 22°C when the substrates started to melt. All experiments were performed with 0.5°C/min heating rate, mimicking the conditions in the optical observations of the emulsions.

The three-phase contact angle measurements at the water-melted oil-frozen oil contact line were performed with $C_{14}TG$ substrates, placed in a glass cuvette. On top of the solid lipid substrate we placed a drop of the liquid MCT oil and, afterwards, we poured gently the studied surfactant solution using a syringe. The three-phase contact angle formed was monitored for 5 min at temperature of 25°C.

All values of the measured three-phase contact angles presented in the manuscript are measured through the aqueous phase.

*Differential scanning calorimetry (DSC)*

The phase behavior of CNO and $C_{12}$TG was studied with DSC on apparatus DSC 250 (TA Instruments, USA). Before measurements, each sample was weighted, placed into a DSC pan (Tzero pan, TA Instruments) and sealed with a hermetic lid (Tzero hermetic lid, TA Instruments) on a Tzero sample press. The samples were cooled and heated at fixed rate, varied between 0.5 and 5°C/min. The DSC curves upon both cooling and heating were recorded. The peak integration of the curves was performed using the build-in functions of the TRIOS data analysis software (TA Instruments).

*Equilibrium surface tension (SFT) and dynamic surface tension (DST)*

The equilibrium surface tensions (SFT) of the surfactant solutions were measured by the Wilhelmy plate method on tensiometer K100 (Krüss GmbH, Germany) at 10°C and 20°C. For some of the systems we measured the surface tensions also by drop shape analysis on DSA30 tensiometer (Krüss GmbH, Germany). The results obtained with the Wilhelmy plate and the drop shape analysis methods were in a very good agreement. The only exception was $C_{12}$SorbEO$_{20}$ + $C_{18:1}$DG solution due to the very slow adsorption kinetics of $C_{18:1}$DG to the surface. For this specific system, the values from the drop shape analysis method are present in the text.

Measurements of dynamic surface tension (DST) of the surfactant solutions were performed at 20°C by the maximum bubble pressure method on tensiometer BP2 (Krüss GmbH, Germany). For some of the systems, we measured the DST also at 10°C. The results obtained at these two temperatures were within the frame of our experimental accuracy (± 0.5 mN/m).

*Freeze-thawing of the lipid drops in bulk emulsion samples*

The drop size evolution in bulk emulsions was studied with either 10 or 20 ml samples, placed in enclosed glass containers. Each freeze-thaw cycle consisted of the following steps: first, the bulk samples were cooled in a freezer for 10-15 min to ensure that an internal temperature of 1-2°C was reached at which the lipid drops, pre-dispersed in the surfactant solution, froze into solid lipid particles. Next, the glass containers were transferred into a cryo-thermostat and were gradually heated with a rate of 0.5°C/min up to several degrees above the melting point of the oil. No freezing of the aqueous phase was allowed in these experiments. After each freeze-thaw cycle for the lipid particles, an aliquot of 1 ml emulsion was separated and stored at room temperature. The obtained disintegrated specimens were first studied under optical microscope to determine qualitatively whether micrometer-sized droplets had remained in the samples. The exact drop size distribution was then measured via DLS.

*Dynamic light scattering (DLS) measurements of drop-size distribution and mean drop size*

The mean drop size after one to several consecutive freeze-thaw cycles was determined by DLS measurements on instrument 4700C (Malvern Instruments, U.K.), equipped with a solid state laser, operating at 514 nm. The built-in multimodal software was used for analysis of the autocorrelation function of the scattered light.

*Purification of commercial monoolein sample of technical grade*

To study the difference between highly purified monoolein (1-oleoyl-rac-glycerol ≥ 99 %, Sigma-Aldrich) and its technical grade analogues (1-oleoyl-rac-glycerol 40 %, Sigma-Aldrich; Peceol, Gattefosse; Aldo MO, Lonza; and Monoolein 40%, TCI Chemicals), we used aqueous alcohol extraction for the 40 % 1-oleoyl-rac-glycerol (Sigma Aldrich). For convenience, the latter substance is denoted hereafter as $GMO_{40}$. The procedure used for $GMO_{40}$ purification was suggested by Sanchez *et. al.*[41] and we slightly modified it to fit better our purpose.

Briefly, we prepared 10 wt. % dispersion of $GMO_{40}$ in aqueous ethanol solution, containing 65 wt. % ethanol and 35 wt. % water (65 wt. % aqueous ethanol). The $GMO_{40}$ dispersion was stirred intensively at 5°C for 2 h. Then it was centrifuged for 1 h at 5°C and at centrifugation speed of 5000 rpm. After centrifugation, the sediment which contained predominantly the diglyceride fraction was dispersed again into 65 wt. % aqueous ethanol and the described procedure was repeated twice at higher temperature of 15°C. The final solid lipid fraction separated after this procedure, contained ≈ 80 % DG as proven by GC analysis.

The upper phase obtained after the initial centrifugation, contained predominantly the monoglyceride fraction. After centrifugation, it was carefully decanted and placed into another bottle. Then the ethanol-water mixture was evaporated fully for 2 h at 50°C. The obtained solid phase was re-dissolved in 95 wt. % ethanol solution, stirred for 2 h at 5°C, and centrifuged at 15°C for 1 h. Then the upper phase was again separated and the ethanol was evaporated from it at 50°C. Thus obtained lipid fraction contained ≥ 90 % monoglycerides.

*Gas chromatography (GC) analysis*

To analyze the commercial oils and the fractions prepared after $GMO_{40}$ purification, we used GC analysis with both hydrolyzed and non-hydrolyzed samples. The analysis of non-hydrolyzed samples gives information about the MG, DG and TG content of the samples, whereas the analysis of the hydrolyzed samples gives information about the chain length distribution of all these components.

Non-hydrolyzed samples were dissolved in chloroform and analyzed by 8890 GC System, Agilent, USA. Comparison between the retardation time for standards (highly purified alkylglycerides) and the analyzed sample was used for peaks' identification.

The samples were hydrolyzed by a procedure, adapted from IUPAC standard methods for analysis.[42,43] The studied sample was dissolved at 5 wt. % in 3.33 M KOH solution in 80 % aqueous ethanol. This solution was stored for 4 h at 40°C and shaken every hour. Afterwards, the sample was dried in a vacuum drier and, then, dissolved in 0.37 M HCl. Next, chloroform was added for extraction and the sample was sonicated in a bath for 15 min. Afterwards, the sample was centrifuged for 30 min to separate well the water (top) and chloroform (bottom) phases. The chloroform phase was collected with syringe and analyzed with GC.

**RESULTS AND DISCUSSION**

In this section we present our experimental results about the cold-burst phenomenon. The systematic experiments are performed with CNO and are presented first. We discuss the obtained results in view of their mechanistic understanding and clarify the main physicochemical factors which control the cold-burst process and can be used for its optimization. Afterwards, results with other tri- and di-glyceride oil mixtures are presented to illustrate the wide range of practically important lipid substances to which the cold-burst method could be applied.

1. **CNO cold-bursting with $C_{12}SorbEO_{20}$ and $C_{18:1}MG$ surfactants and their mixtures**

Up to now, no studies have been published to clarify whether nanovoids are formed upon crystallization of CNO, whereas such process was reported for cocoa butter.[44] As explained in Ref. 7, these nanovoids can be flooded by the aqueous phase when appropriate surfactant is used. Depending on the three-phase contact angle, θ, formed at the three-phase contact line air-surfactant solution-solid lipid substrate, the behavior of the dispersed lipid particles upon storage/melting is different. Intensive particle disintegration is observed when the contact angle is smaller than *ca.* 50-60°, whereas double water-in oil-in water emulsion droplets form in the moment of lipid melting when this angle is higher than *ca.* 90-100°. The role of the angle θ and the used method for its measurement are schematically illustrated in Figure 3a,b.

The most significant difference between the model triglycerides, studied in our previous paper,[7] and CNO is that the TGs containing only one type of alkyl chains (e.g. trimyristine) remain solid throughout the entire heating period until the final melting temperature is reached. In contrast, CNO contains a mixture of various triglycerides, some of which melt at much lower temperature compared to the others which contain longer saturated chains, see the heating DSC thermograms presented in Figure 2. Therefore, upon heating, some regions inside the frozen

CNO particles melt before the others, thus forming co-existing liquid and frozen regions inside these particles, in a certain temperature range below the complete particle melting.

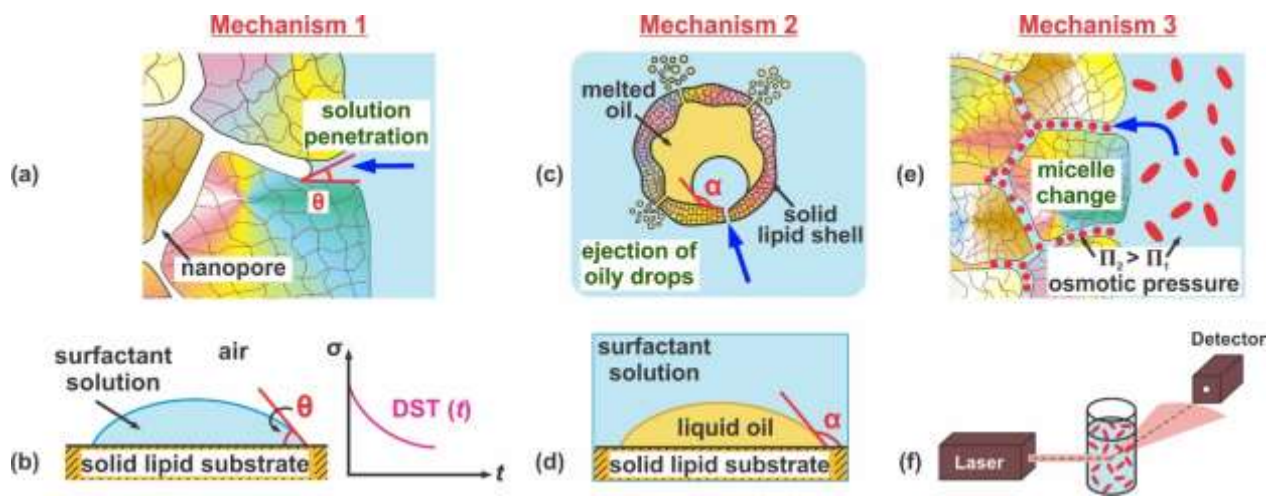

**Figure 3. Main "elementary" mechanisms driving the water influx into the lipid particles in the cold-burst process.** (a) Wetting of the nanopores by the aqueous surfactant solution, governed by the three-phase contact angle, $\theta$, formed at the contact line air-surfactant solution-solid lipid; (b) Model experiment used to measure $\theta$. In the current study we found that the kinetics of surfactant adsorption $\sigma(t)$ is also important to ensure the rapid decrease of $\theta$; (c) Wetting of the internal surface of the solid lipid shell in a partially melted lipid globule, governed by the three-phase contact angle, $\alpha$, formed at the contact line surfactant solution-liquid oil-frozen lipid; (d) Model experiment used to measure $\alpha$; (e) Change in the size of the surfactant aggregates upon their penetration through the nanopores with the formation of higher number of smaller micelles which create higher osmotic pressure than the original surfactant solution; (f) Dynamic light scattering was used to determine the size of the surfactant micelles. Most efficient particle disintegration was observed when all three mechanisms were operative. The schematics are not in scale and illustrate the main processes only.

As explained below in relation to the newly obtained results with CNO and the other oils studied in the current article, this wide range of melting temperatures leads to the appearance of two additional mechanisms (Figures 3c and 3e) which have not been reported in the experiments with single triglycerides, described in Ref. 7. These mechanisms are discussed in the following sections to explain the microscope observations with the various systems studied. As explained below, these additional mechanisms can be realized only if the requirement for the mechanism shown in Figure 3a (viz. wetting of the nanopore surface by the surfactant solution) is already fulfilled. In other words, the new mechanisms complement the initial one and, when present, make the cold-burst process much more efficient for CNO and for the other natural oils. The experimental methods, illustrated schematically in Figures 3d and 3f, were used to prove the importance of the additional two mechanisms, as explained below.

In our previous study with pure triglycerides, we showed that most efficient self-dispersion is observed when water-soluble and oil-soluble surfactants are combined in the aqueous phase.[7] Therefore, in the current study we chose one water-soluble surfactant allowed to be used in foods, $C_{12}SorbEO_{20}$ (trade name Tween 20), and performed a series of systematic experiments, combining it with various oil-soluble surfactants, see Supplementary Table S3. For reasons explained in the next section, we observed most efficient cold-burst process with 99 % $C_{18:1}MG$ as oil-soluble surfactant, see Figure 4 and Supplementary Movies 2 and 3.

*Optical observations of CNO particles in solutions of $C_{12}SorbEO_{20}$ and $C_{18:1}MG$*

In the absence of oil-soluble surfactant, the frozen CNO particles dispersed in $C_{12}SorbEO_{20}$ solution melted directly, returning to the initial emulsion drops, Figure 4a. When $C_{18:1}MG$ surfactant was present in the aqueous solution, the particle behavior was very different. In these experiments, we usually cooled the sample quickly from 25°C to 2°C with cooling rate of *ca.* 2.5°C/s and then started the heating with 0.5°C/min. Depending on the initial drop size, the frozen particles began to disintegrate at temperature around $12 \pm 2$°C, with numerous small lipid particles detaching from the periphery of each original big particle. The process became more intensive with the increase of temperature, see Figure 4b,c. In the experiments with ≤ 20 μm drops, the particles disintegrated completely at 18-19°C into numerous, much smaller particles, due to the penetration of the aqueous phase, Figure 4b. These observations correspond to the combination of mechanisms 1 and 3, as illustrated in Figure 3a and 3e.

With bigger in size CNO particles (diameter > 20 μm), two simultaneous processes were observed, see Supplementary Movie 3 and Figure 4c. Together with the detachment of the small particles from the particle periphery, described above, we observed also a phase separation of the TG fraction with lower melting temperature from the TG fraction with higher melting temperature (*i.e.* of the TG with shorter and/or unsaturated chains from those with saturated and longer alkyl chains). The low-melting-temperature TGs spontaneously left the interior of the frozen particle into the aqueous solution in the form of small liquid droplets at temperatures which were much lower than the melting temperature of the bulk CNO, see Figure 4c. At higher temperatures, but still lower than the melting temperature of the bulk CNO, bursting of the remaining solid shell of the high-temperature-melting fraction of CNO was observed, due to intensive penetration of the aqueous phase into this shell. In this case, the combination of all three mechanisms is observed – see, *e.g.*, the ejected droplets shown by white arrows in Figure 4c which correspond to Mechanism 2 in Figure 3.

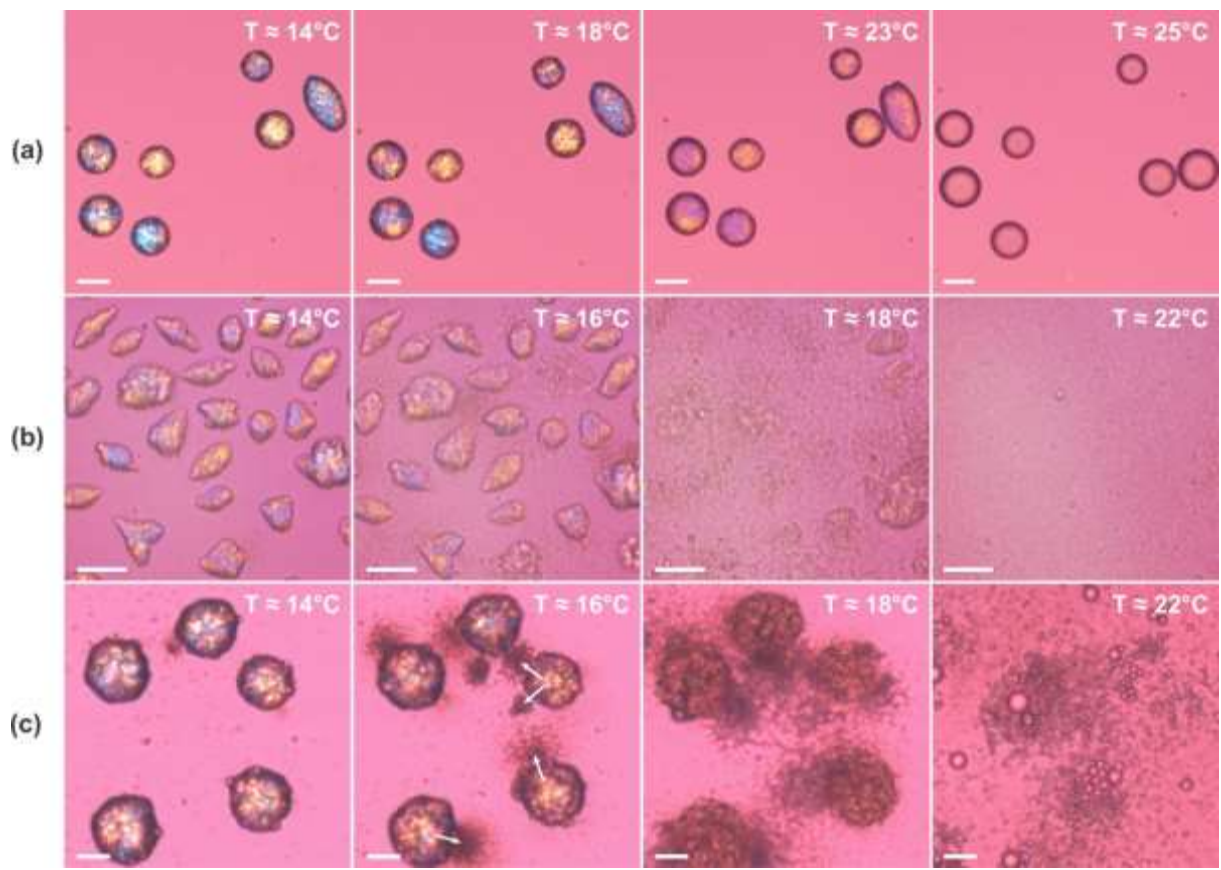

**Figure 4. Optical microscopy experiments with CNO drops, dispersed in 1.5 wt. % $C_{12}SorbEO_{20}$ in absence (a) and in presence of 0.5 wt. % $C_{18:1}MG$ (b and c).** (a) Particles stabilized by $C_{12}SorbEO_{20}$ surfactant only do not disintegrate upon heating. (b) 15 μm particles disintegrate completely upon heating. (c) Disintegration with 33 μm particles proceeds via two coupled mechanisms: initially, small liquid droplets are ejected from the particle interior (see the white arrows at 16°C); upon further heating, the solid shell of the particles disintegrates completely. The size of the drops is reduced significantly, but some micrometer sized drops remain after the particle complete melting. All experiments are performed with heating rate of 0.5°C/min. Scale bars in all images, 20 μm.

It is worth noting that the first process of separation of the oily drops from the solid fraction of CNO was very similar in nature to the dewetting process described previously with lipid globules composed of mixed soybean oil (low-melting-temperature) and tristearin (high-melting temperature), dispersed in aqueous surfactant solutions.[5] In these latter experiments, a release of the liquid soybean oil from the frozen network of tristearin microsrystals was observed by optical microscopy, when appropriate surfactants were used: LAS + EO7 (linear alkylbenzene sulfonate + polyoxyethylene alkyl ether). This dewetting process was driven by the appropriate contact angle of the already melted oil droplets over the (still) frozen lipid substrate. Similar processes of spontaneous separation (dewetting) of the liquid and solid lipid components were observed also with globules composed of alkane mixtures.[45] In the latter study we showed also that the multicomponent alkane particles melted from inside out, i.e. upon freezing, the shorter alkanes were trapped predominantly in the particle interior, whereas the longer-chain alkanes formed a solid shell with higher melting temperature at the particle periphery – a structure which we saw in the current experiments with large CNO particles as well.

*Effect of $C_{12}SorbEO_{20}$ and $C_{18:1}MG$ total concentration*

Most of the experiments in this study were performed at 2 wt. % total surfactant concentration and 3:1 ratio of the water- to-oil-soluble surfactants, because this concentration is sufficiently high to ensure a complete coverage of the surface of the newly formed small particles/droplets (up to *ca.* 3 vol. % of oil in the emulsion). This concentration ensures also stable surfactant solutions which do not phase-separate upon prolonged shelf-storage.

To clarify the effect of surfactant concentration, we performed model experiments also with 1 wt. % and 0.5 wt. % total surfactant concentration, at the same 3:1 surfactant ratio. With 1 wt. %, the self-dispersion efficiency was almost unchanged, compared to the one with 2 wt. %. At 0.5 wt. % the cold-bursting process was still observed with some small droplets separating from the frozen particle periphery. However, the final drop size remained much bigger in the latter experiment compared with the experiments at higher surfactant concentrations. At surfactant concentrations lower than 0.5 wt. %, the cold-burst process was not observed. On the other hand, at total surfactant concentrations > 2 wt. %, the cold-bursting process was observed until the aqueous solution became gelled (around ca. 6 wt. %), due to the entanglement of the thread-like micelles formed in such highly concentrated surfactant solutions.

*Effects of the ratio $C_{12}SorbEO_{20}$ to $C_{18:1}MG$, and of the phase in which $C_{18:1}MG$ is dissolved initially*

We observed efficient cold-bursting with $C_{12}SorbEO_{20}$ when $C_{18:1}MG$ was dispersed in the aqueous phase or in both the aqueous and the oily phases. In contrast, no particle bursting was observed when the oil-soluble surfactant was pre-dissolved in the oily phase only. This is related to the ability for realization of Mechanism 3 as shown schematically in Figure 3e.

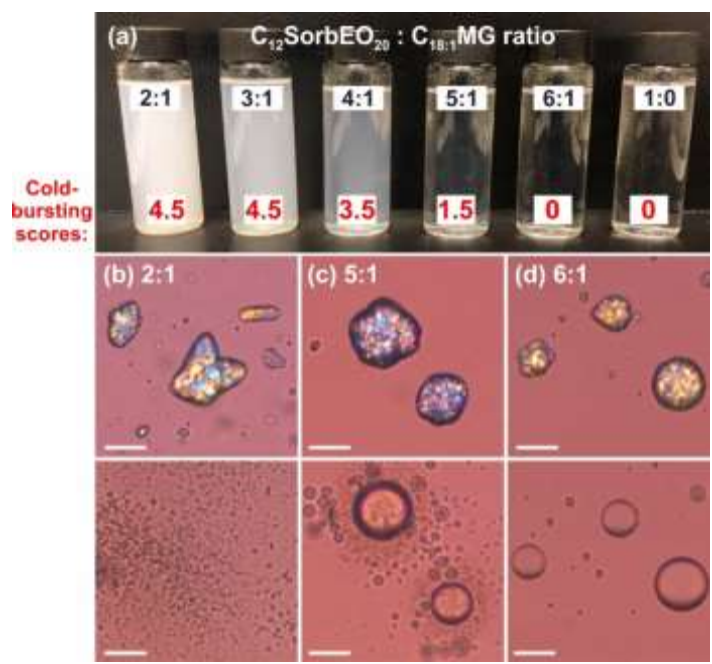

**Figure 5. Effect of surfactant ratio on the efficiency of the cold-bursting process.** (a) Images of glass bottles containing aqueous surfactant solutions of 1.5 wt. % $C_{12}SorbEO_{20}$ and $C_{18:1}MG$, mixed in different weight ratios, as indicated on the images (no CNO is present in these bottles). The cold-bursting scores given with red numbers on these images refer to the efficiency of the bursting process observed with CNO particles in the respective surfactant solution: "0" no particle disintegration observed; "1" small droplets are separated from the surface of the frozen particle, but the mean drop size does not change significantly; "2-3" all initial drops disintegrate into smaller micrometer drops; "4-5" most of the drops disintegrate into much smaller droplets with size < 1 μm, but very small fraction of micrometer drops may remain in the sample. (b-d) The images in the upper row show frozen CNO particles, dispersed in the surfactant solutions with the surfactant ratio indicated. The bottom row shows images of the particles in the same samples, after one freeze-thaw cycle of CNO: in (b,c) the drop size decreases significantly in the emulsions stabilized with the opalescent 3:1 and the turbid 2:1 surfactant mixtures, while in (d) no drop disintegration is observed with the transparent 6:1 solution. The mechanistic explanation of this observation is given later in the main text. The heating rate in all these experiments was 0.5°C/min. Scale bars, 20 μm.

The proportion of $C_{12}SorbEO_{20}$ to $C_{18:1}MG$ was also very important. As seen from Figure 5, depending on the ratio between the water-soluble and oil-soluble surfactants, the aqueous surfactant solutions appeared differently (no CNO was added in the samples shown in Figure 5a). The surfactant solutions were completely transparent when the ratio between the water-soluble surfactant, $C_{12}SorbEO_{20}$, and oil-soluble surfactant, $C_{18:1}MG$, was 6:1 or higher. The surfactant solutions were almost transparent at 5:1 ratio, opalescent at 4:1 and 3:1 ratios, and became very turbid at lower ratio, *viz.* at higher fraction of the oil-soluble surfactant, Figure 5.

These observations could be explained by the molecular surfactant aggregates formed when the oil-soluble surfactant is introduced into the aqueous solution. When the water-soluble surfactant is in a big excess, the oil-soluble surfactant is completely solubilized in the mixed surfactant micelles. These surfactant solutions are transparent, because the size of the micelles dominated by water-soluble surfactants is typically much smaller than the wave-length of the visible light, ca. 5-10 nm.[46] In contrast, when the fraction of oil-soluble surfactant approaches or exceeds the solubilization capacity of the micelles, much larger, non-spherical mixed molecular aggregates, vesicles and/or micellar agglomerates are formed which have much bigger volume and higher light scattering power than the spherical micelles of the water-soluble surfactants. As a result, the surfactant solution becomes opalescent or turbid and the oil-soluble surfactant can phase-separate upon prolonged shelf-storage.

Our experiments with this surfactants showed that cold-bursting was observed with all solutions which were opalescent or turbid. In contrast, the cold-burst process was suppressed with the optically clear surfactant solutions. For example, very efficient cold-bursting was observed at 2:1 and 3:1 ratios of $C_{12}SorbEO_{20}$ with $C_{18:1}MG$ (Figure 5b), whereas the cold-bursting efficiency decreased for 4:1 and 5:1 ratios (Figure 5c) and was almost completely suppressed at 6:1 ratio (Figure 5d).

It is worth noting that, when CNO emulsion was prepared in $C_{12}SorbEO_{20}$ + $C_{18:1}MG$ mixed aqueous solution and was then stored at temperatures above the CNO melting temperature, some fraction of the oil-soluble surfactant $C_{18:1}MG$ gradually transferred from the aqueous phase into the oily drops. As a result, the aqueous surfactant solution became completely transparent (checked after the oil drops creamed due to buoyancy). When this stored emulsion was examined, no cold-bursting was observed, because the surfactant ratio in the aqueous phase was shifted to lower oil-soluble/water-soluble ratio. However, cold-bursting was observed again when the same CNO drops were dispersed in freshly prepared aqueous solution of $C_{12}SorbEO_{20}$ + $C_{18:1}MG$ mixture of appropriate ratio.

These latter results prove that it is essential for the cold-burst process to have mixed micelles with appropriate ratio of $C_{12}SorbEO_{20}$ and $C_{18:1}MG$ in the aqueous phase (does not matter whether oil-soluble surfactant is present in the oily phase). This important observation is related to the possibility for realization of Mechanism 3 in Figure 3e, as explained in details in the Mechanism section below.

*Effect of glyceryl dioleate*

The glyceryl monooleate surfactant of high purity (> 99 %) is rather expensive. Therefore, knowing that this surfactant induce very efficient cold-burst process, we checked the effect of its analogues of technical grade. According to their producers, the monooleins of technical grade usually contain > 40 % $C_{18:1}MG$. We tested four different monoolein samples from different producers, but surprisingly the cold-burst process was not observed with neither of them at the standard 3:1 water- to oil-soluble surfactant ratio and 2 wt. % total surfactant concentration although the surfactant solutions were turbid, see Supplementary Figure S1.

The main impurity in the technical monooleins are the glyceryl dioleates ($C_{18:1}DG$), therefore we hypothesized that their presence interrupts the cold-burst process by affecting the relevant interfacial (surface) tensions which in turn control the wetting ability of the surfactant solution, interrupting the realization of Mechanism 1 in Figure 3 (see also section Mechanism below).

To obtain $C_{18:1}DG$, we purified one of the commercial monooleins of technical grade (1-oleoyl-rac-glycerol > 40 %, Sigma-Aldrich, denoted as $GMO_{40}$) as described in Methods section. Our GC analysis showed that the initial $GMO_{40}$ source contained ca. 65 % MG, 33 % DG and 2 % TG, with $C_{18:1}$ chains being the predominant fraction ≈ 85 %. After the purification procedure we obtained separate fractions of $C_{18:1}MG$ (purity ≥ 90 %) and $C_{18:1}DG$ (purity ≈ 80 %).

Microscopy observations with emulsions stabilized by surfactant mixtures containing these fractions confirmed, as expected, that the cold-burst process is observed with the $C_{18:1}MG$ fraction, whereas it was not observed with $C_{18:1}DG$ fraction. No significant differences were observed between the behavior of the emulsions, prepared with the $C_{18:1}MG$ fraction obtained after $GMO_{40}$ purification, and that of the emulsions prepared with 99% pure $C_{18:1}MG$ purchased.

We also checked whether the absence of cold-burst process in samples with $GMO_{40}$ is due to the lower concentration of $C_{18:1}MG$ in these experiments compared to the case when pure $C_{18:1}MG$ is added at 0.5 wt. % concentration. Experiments with 1.5 wt. % $C_{12}SorbEO_{20}$ and higher concentrations of $GMO_{40}$ showed that this is not the explanation, viz. no disintegration was observed for CNO dispersed in 1.5 wt. % $C_{12}SorbEO_{20}$ + 0.8 wt. % $GMO_{40}$ solution.

The reason for the absence of disintegration in the samples with $GMO_{40}$ surfactant turned out to be that in presence of $C_{18:1}DG$, the molecules of $C_{18:1}MG$ are included in mixed micelles, which significantly lowers their surface activity. In turn, this leads to elevated values of the surface tension and of the three phase contact angle at the CNO-air-solution interface for $C_{12}SorbEO_{20}$ + $GMO_{40}$ compared to the $C_{12}SorbEO_{20}$ + $C_{18:1}MG$ solution, see Supplementary Table S3, Figure 6 and the explanations of the cold-bursting mechanism in the following section.

*Mechanism of the process*

A good correlation between the three phase contact angle formed at the air-solution-lipid interface and the cold-bursting phenomenon was found in our previous study.[7] An intensive cold-bursting was observed when the three phase contact angle was relatively low, *ca.* < 50-60°. Such smaller angles favor the wetting of the surface of the solid triglyceride by the surfactant solution, thus promoting the penetration of the aqueous phase into the porous network inside the frozen lipid particle, Mechanism 1 in Figure 3.

To clarify the mechanism of CNO cold-bursting observed with the combinations of oil-soluble and water-soluble surfactants, we performed similar contact angle measurements. An aqueous surfactant solution drop was placed on top of a frozen CNO substrate and we measured the three-phase contact angle dependence as a function of temperature, see Figure 6 for illustrative results and Figure 3b for schematic presentation of the experiment.

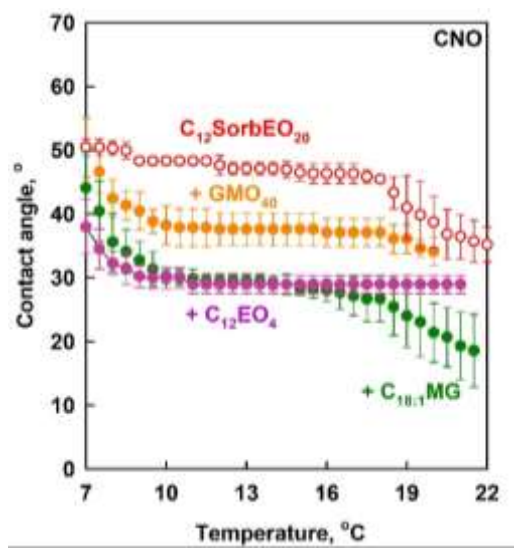

**Figure 6. Temperature dependence of the contact angles, measured with drops of aqueous surfactant solutions on frozen CNO substrate.** All solutions contained 1.5 wt. % $C_{12}SorbEO_{20}$. The empty circles show the results obtained with solutions of this water-soluble surfactant only, whereas the full circles show results from measurements in the presence of additional 0.5 wt. % oil-soluble surfactant: $C_{18:1}MG$ (green); $C_{12}EO_4$ (purple); $GMO_{40}$ (orange).

These experiments showed that the contact angle of drops of deionized water, placed on CNO solid substrate, is $\theta = 83 \pm 1°$ (at 10°C) and slightly decreases when the temperature is increased above 18-19°C, due to the formation of heterogenic structure on the CNO substrate at these temperatures, see the DSC thermogram for bulk CNO shown in Figure 2b. When the same measurements were performed with 1.5 wt. % $C_{12}SorbEO_{20}$ aqueous solution, the three phase contact angle was significantly lower, $\theta \approx 48 \pm 1°$ at 10°C, empty red circles in Figure 6. The contact angle decreased to $30 \pm 1°$ in the presence of 0.5 wt. % $C_{18:1}MG$. With this mixed surfactant system, very intensive particle disintegration was observed, as explained above. For the system 1.5 wt. % $C_{12}SorbEO_{20}$ + 0.25 wt. % $C_{18:1}MG$, we measured contact angle of $40 \pm 1°$ at 10°C. This result is in a good agreement with the observation that the cold-bursting was significantly suppressed at 0.25 wt. % $C_{18:1}$MG (ratio 6:1) compared with the bursting observed with 0.5 wt. % $C_{18:1}$MG (ratio 3:1).

The observation that the self-dispersion process is efficient only when the $C_{18:1}$MG surfactant was dispersed in the aqueous phase or in both the aqueous and oily phases, is also in agreement with the contact angle measurements. Similar contact angles were measured for $C_{12}SorbEO_{20}$ on CNO and CNO + 0.5 wt. % $C_{18:1}$MG substrates (the monoglyceride was dissolved in the CNO before its freezing to prepare the solid substrate), see the empty red circles in Figure 6 and Supplementary Figure S2. When $C_{18:1}$MG was present in the aqueous phase, the contact angle was significantly lower, see the full green circles in Figure 6 and Supplementary Figure S2.

From the experiments performed at different ratios of water-to-oil-soluble surfactants, we concluded that we need an excess fraction of the oil-soluble surfactant in the aqueous phase, forming relatively large surfactant aggregates (opalescent solutions). When such solution penetrates inside the porous network of the frozen lipid particle, the oil-soluble surfactant preferentially adsorbs on the surface of the lipid nanopores. As a result, the penetrating micelles get enriched in water-soluble surfactant which leads to shape transformation into spherical micelles which are smaller in size (ca. 6-10 nm in diameter) and much larger in number concentration. The difference in the micellar number concentration inside the porous network of the lipid particle and in the surrounding them main aqueous phase causes a strong osmotic effect which sucks water into the pores, see Figure 3e for schematic presentation of this mechanism. As a result of this osmotic effect, the aqueous phase floods the particle interior, increasing the particle volume and tearing apart the individual crystalline domains. Note that the osmotic effect is efficient only if the aqueous phase is able to penetrate the porous network in the frozen lipid particle which, in turn, is controlled by the three-phase contact angle, θ, discussed above.

From the experiments with $C_{12}SorbEO_{20}$ and $C_{18:1}$MG we can conclude that the following requirements should be fulfilled for efficient particle self-disintegration:

(1) The three phase contact angle at the CNO-water-air interface should be lower than *ca.* 35°.
(2) The micelles in the surfactant solution should reduce their aggregation number (size) upon adsorption of the oil-soluble surfactant on the surface of the lipid nanopores. This change in micelle aggregation number causes an osmotic pressure difference which sucks the aqueous phase into the frozen particles, leading to their disintegration.

The described mechanism explains also why no drop disintegration is observed for $C_{12}SorbEO_{20}$ + $GMO_{40}$ and $C_{12}SorbEO_{20}$ + $C_{18:1}$DG solutions. As seen in Supplementary Table

S3, the contact angle and the surface tension measured with $C_{12}SorbEO_{20}$ solutions is practically unaffected by the addition of $C_{18:1}DG$: $\theta_{CNO} \approx 48°C$ and $\sigma_{aw} \approx 37.7$ mN/m at 10°C. The latter comparison shows that the surfactant adsorption layer of $C_{12}SorbEO_{20}$ remains unaffected by the presence of glyceryl dioleates. The angle measured for $C_{12}SorbEO_{20} + GMO_{40}$ is somewhat lower, *ca.* 38° under equivalent conditions, however, this value is still much higher than the one measured for $C_{12}SorbEO_{20} + C_{18:1}MG$, $\theta_{CNO} \approx 30°$ at 10°C. From these results, we conclude that the presence of glyceryl dioleates in the commercial grade monooleins leads to the formation of mixed $C_{18:1}MG + C_{18:1}DG$ micelles which suppress the release of $C_{18:1}MG$ molecules, needed to effectuate the cold-burst process in this system.

*Effect of the cooling and heating rates*

To clarify the effect of the cooling and heating rates for the particle disintegration, we performed series of experiments with the same initial monodisperse emulsion ($d_{ini} \approx 30$ μm), varying the temperature protocol. Figure 7a shows microscope images of the initial droplets, whereas Figure 7b-e shows the droplets obtained after one cooling-heating cycle at different rates of cooling and heating, as indicated on the images.

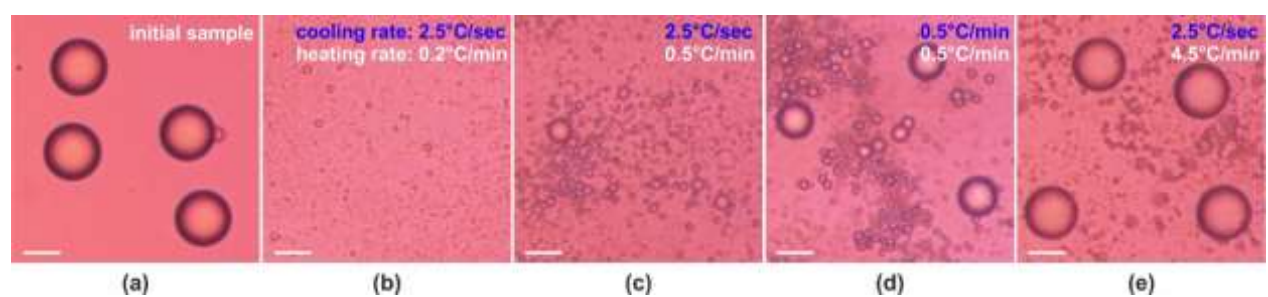

**Figure 7. Images illustrating the strong dependence of the cold-burst process on the cooling and heating rates.** These experiments are performed with monodisperse emulsion of initial drop size ≈ 30 μm. Most efficient drop size decrease is observed when fast heating-slow cooling protocol is applied. The cooling rates are given with blue labels and the heating rates – with white labels on the images. Scale bars, 20 μm. All drops are dispersed in 1.5 wt. % $C_{12}SorbEO_{20}$ + 0.5 wt. % $C_{18:1}MG$ surfactant solution.

The drop size decreases most significantly when rapid cooling and slow heating protocol is applied, Figure 7b. The rapid cooling ensures formation of numerous nuclei and small crystalline domains in the frozen lipid particles. The slow heating ensures sufficient time for the aqueous phase to penetrate deep into the particle interior and to burst the particles. Note the significant difference between 0.5°C/min and 0.2°C/min heating rates, cf. Figure 7b and 7c. At 0.2°C/min heating, all drops disintegrate into thousands of sub-micrometer droplets and single 2-3 μm drops remain in the sample, whereas at 0.5°C/min many micrometer droplets remain. The comparison of the samples obtained after rapid or slow cooling, at fixed heating rate of 0.5°C/min (cf. Figure 7c and 7d), shows that much smaller drops are observed after fast cooling.

We conclude that the most efficient cold-burst protocol is rapid cooling and slow heating. Therefore, all other experiments were performed with cooling rate of ≈ 2.5°C/s and heating rate of 0.5°C/min, unless otherwise specified.

*Minimum achievable drop size*

To obtain smaller CNO particles, we performed several consecutive cooling-heating cycles with 1 wt. % CNO emulsion, containing initially 5 μm drops, dispersed 1.5 wt. % $C_{12}SorbEO_{20}$ + 0.5 wt. % $C_{18:1}MG$ solution. Illustrative microscopy images of the initial sample and of the emulsified sample after one and three consecutive cooling-heating cycles are presented in Supplementary Figure S3. After the first cycle, the CNO drops had volume-average diameter, $d_V$ ≈ 580 ± 80 nm, and number-average diameter, $d_N$ ≈ 320 ± 35 nm. After the second cooling-heating cycle, $d_V$ decreased down to ≈ 470 ± 25 nm. After three consecutive cooling-heating cycles, the measured drop sizes were $d_V$ ≈ 320 ± 15 nm and $d_N$ ≈ 170 ± 40 nm.

As explained in the previous section, the cooling-heating protocol has strong impact on the drop size distribution obtained. Therefore if the protocol is optimized even further (*e.g.* using even slower heating rates) one can expect to obtain even smaller droplets as demonstrated with the other mixed triglyceride compounds in Section 3 below.

Very similar results about the mean drop diameters were obtained also with 5 wt. % CNO emulsions in which the surfactant concentration was tripled, i.e. 4.5 wt. % $C_{12}SorbEO_{20}$ + 1.5 wt. % $C_{18:1}MG$, which shows that these emulsions contained sufficient surfactant to cover the surface of the newly formed particles and droplets.

## 2. **Effect of different surfactants for the CNO cold-bursting**

After we clarified the cold-bursting mechanisms for CNO particles in $C_{12}SorbEO_{20}$ + $C_{18:1}MG$ solutions, we performed series of experiment with various surfactant combinations to validate our conclusions and to expand our understanding. First, we present the results obtained with the same hydrophilic surfactant, $C_{12}SorbEO_{20}$, and a series of oil-soluble surfactants for comparison. Next, we present results with the same hydrophobic surfactant, $C_{18:1}MG$, in combination with a series of nonionic water-soluble surfactants. Finally, we show that the same approach and conclusions can be applied also to ionic surfactants. Thus, we demonstrate that the cold-burst process is observed with very wide range of systems when the appropriate conditions are met.

*Different oil-soluble surfactants in combination with $C_{12}SorbEO_{20}$*

As explained in section Mechanism above, the addition of oil-soluble surfactant into the aqueous solution decreases the contact angle, $\theta$, compared to the one measured with $C_{12}SorbEO_{20}$ only. This decrease depends on the specific oil-soluble surfactant. The angle drops by *ca.* 5° when $C_{18:1}EO_2$ (see Supplementary Table S3) or $C_{12}Sorb$ is added to $C_{12}SorbEO_{20}$. However, this decrease was not sufficient to induce an intensive cold-bursting.

The contact angle decreased more significantly when $C_{12}EO_4$ or $C_{18:1}MG$ were added into the aqueous phase, Figure 6. However, for the system 1.5 wt. % $C_{12}SorbEO_{20}$ + 0.5 wt. % $C_{12}EO_4$ we observed very limited disintegration, see Supplementary Figure S4. This surfactant is with intermediate hydrophobic chain of 12 carbon atoms, which is exactly the same as that of the main water-soluble surfactant. As a result, spherical mixed micelles are formed in the presence of $C_{12}EO_4$ molecules and the respective surfactant solution is completely transparent. Therefore, no osmotic effects are present (Mechanism 3, Figure 3) in the mixed solution 1.5 wt. % $C_{12}SorbEO_{20}$ + 0.5 wt. % $C_{12}EO_4$ and the cold-burst process is non-intensive. To confirm that the absence of osmotic effects is the main problem with this surfactant system, we doubled the $C_{12}EO_4$ concentration in the mixed surfactant solution. Indeed, the solution of 1.5 wt. % $C_{12}SorbEO_{20}$ + 1 wt. % $C_{12}EO_4$ was opalescent and, as expected, the CNO particles disintegrated completely after one freeze-thaw cycle, see Supplementary Figure S4.

Similar effect was observed with the $C_{12}SorbEO_{20}$ + $C_{10}MG$ solutions. For this system, the measured contact angle was even lower than the one for $C_{12}EO_4$ and $C_{18:1}MG$, ca. 26 ± 2°. However, at 0.5 wt. % $C_{10}MG$, the surfactant solution was completely clear, limiting the osmotic pressure effect. When we doubled the $C_{10}MG$ concentration to 1 wt. %, we obtained opalescent surfactant solution and observed much more intensive cold-bursting process.

Therefore, from these experiments we confirmed that both the contact angle, $\theta$, and the type of surfactant aggregates are crucial to induce an intensive cold-burst process, viz. both Mechanisms 1 and 3 illustrated in Figure 3 should be operative. The osmotic effect could be ensured using appropriate ratio of the oil-soluble to water-soluble surfactants when bigger molecular aggregates are formed and the mixed surfactant solution is opalescent.

*Different water-soluble surfactants in combination with $C_{18:1}MG$ and $C_{12}EO_4$*

Similar contact angles were measured with 1.5 wt. % of the water-soluble surfactants $C_{16}SorbEO_{20}$, $C_{12}EO_{23}$ or $C_{18}EO_{20}$, when they were combined with 0.5 wt. % $C_{18:1}MG$ – see Supplementary Figure S2 for $C_{16}SorbEO_{20}$, the other curves are completely identical (not shown). At heating rate of 0.5°C/min, an intensive cold-bursting was observed with $C_{12}EO_{23}$. The process was much less pronounced for $C_{18}EO_{20}$ and it was very limited only from the surface for $C_{16}SorbEO_{20}$, despite the low contact angle measured with all these surfactants.

In another series of experiments, we changed the oil-soluble surfactant from $C_{18:1}MG$ to $C_{12}EO_4$ for which we also observed low values of the equilibrium contact angle, θ, see Figure 6 and Supplementary Figure S2. The measured angles for these three systems were identical, see Supplementary Table S3. On the other hand, we observed cold-bursting with surfactants $C_{12}SorbEO_{20}$ and $C_{12}EO_{23}$, but not with $C_{18}EO_{20}$ at the heating rate of 0.5°C/min.

The comparison of these various surfactant systems indicates that some additional factor should be taken into account when the surfactant mixture contains long-chain water-soluble surfactants, e.g. $C_{16}SorbEO_{20}$+ $C_{18:1}MG$ or $C_{18}EO_{20}$+ $C_{12}EO_4$. To check whether some kinetic effects are involved, due to the long chain of the water-soluble surfactant, we measured the dynamic surface tension of these systems, Supplementary Figures S5a and S7. The comparison of the results obtained in the presence and in the absence of $C_{18:1}MG$ (empty and full symbols in Supplementary Figure S7) shows that $C_{18:1}MG$ starts to affect σ(*t*) only at ≈ 800 ms after the surface formation. Also, the initial value of σ(*t*) at short times, measured with the long-chain surfactant $C_{16}SorbEO_{20}$, is by ≈ 7 mN/m higher than that measured with $C_{12}SorbEO_{20}$. Thus we see that the combination of the relatively high σ(*t*) for $C_{16}SorbEO_{20}$ with the slow adsorption of $C_{18:1}MG$ could explain the suppressed particle disintegration in this surfactant system. The measured σ(*t*) for the mixture $C_{18}EO_{20}$ + $C_{18:1}MG$ is similar to that for $C_{16}SorbEO_{20}$ + $C_{18:1}MG$ however, significant difference is observed between the appearance of two solutions, see Supplementary Figure S6. Therefore, the difference in the cold-bursting scores for these system is due to the different osmotic pressure effects, *viz.* Mechanism 3.

In contrast, the combination of short-chain water-soluble surfactant ($C_{12}$) with another short chain oil-soluble surfactant, $C_{12}EO_4$, gives much lower values of σ(*t*), see Supplementary Figure S5a. As explained above, this surfactant combination leads to an intensive cold-burst process when the surfactant solutions are opalescent and the osmotic pressure effect is also present.

Elucidating this role of the kinetics of surfactant adsorption for the cold-burst process, we hypothesized that we should be able to observe this process if we ensure longer time for water penetration into the lipid particles when working with mixture of the long-chain surfactants, $C_{16}SorbEO_{20}+C_{18:1}MG$, because the equilibrium values of $\sigma_{aw}$ and $\theta$ for this mixture are very close to those measured with the system $C_{12}SorbEO_{20}+C_{18:1}MG$. To check this hypothesis, we performed additional experiment with CNO drops dispersed in $C_{16}SorbEO_{20} + C_{18:1}MG$ solution at slower heating rate of 0.2°C/min. Indeed, in these experiments we did observe particle disintegration, see Supplementary Figure S8. It was not as intensive as in the case with $C_{12}SorbEO_{20} + C_{18:1}MG$, but it was rather significant. Probably, even slower heating would boost the particle disintegration, but we did not try such additional (very time consuming) experiments.

*Cold-bursting of CNO with ionic surfactants*

So far we discussed the behavior of the systems containing nonionic surfactants only. It is well known that the ionic surfactants adsorb even faster than the nonionic ones, because the lifetime of the ionic surfactant micelles is shorter.[47] Therefore, in this section we present our results for the cold-bursting process observed with two ionic surfactants (LAS and SLES) and one zwitterionic surfactant (CAPB) which are commonly used in home and personal care cleaning formulations. Note that the ionic surfactant solutions used in these experiments contain electrolytes of high concentration (30 mM $CaCl_2$ added or NaCl present in the commercial CAPB) which suppress the electrostatic barrier upon adsorption of the charged molecules of the ionic surfactants, thus accelerating the surfactant adsorption.

One of the most commonly used ionic surfactants for detergency and laundry formulations is LAS. It has been shown that there is a specific interaction between $Ca^{2+}$ ions and the headgroups of the LAS molecules, resulting in very low interfacial tensions at the oil-water interface.[48] However at concentrations much higher than the critical micellar concentrations, the solutions of LAS + $Ca^{2+}$ are unstable at 1:1 molar ratio and easily precipitate. To stabilize the LAS solutions against precipitation we added a certain fraction of SLES which is calcium-tolerant surfactant. Therefore, the next series of experiments was performed with LAS + SLES in 3:1 weight ratio, at 1.5 wt. % total surfactant concentration, with and without 30 mM $Ca^{2+}$ present in the aqueous phase, see Figure 8.

The contact angles, $\theta$, measured for both solutions (± $Ca^{2+}$) at the frozen CNO-water-air contact line were very low, 30 ± 3° in the absence and 26 ± 2° in the presence of $Ca^{2+}$ in the temperature range between 10 and 20°C, see Figure 8b. For both solutions, the dynamic surface tension $\sigma(t)$ was much lower than that of $C_{12}SorbEO_{20} + C_{18:1}MG$ solution, see Supplementary Figure S5b. However, in the absence of $Ca^{2+}$ we observed very limited particle disintegration.

The whole CNO particles disintegrated completely into numerous smaller particles in the presence of $Ca^{2+}$ only, see Figure 8a. The cold-bursting efficiency in the latter solution was even higher compared to $C_{12}SorbEO_{20} + C_{18:1}MG$ solution.

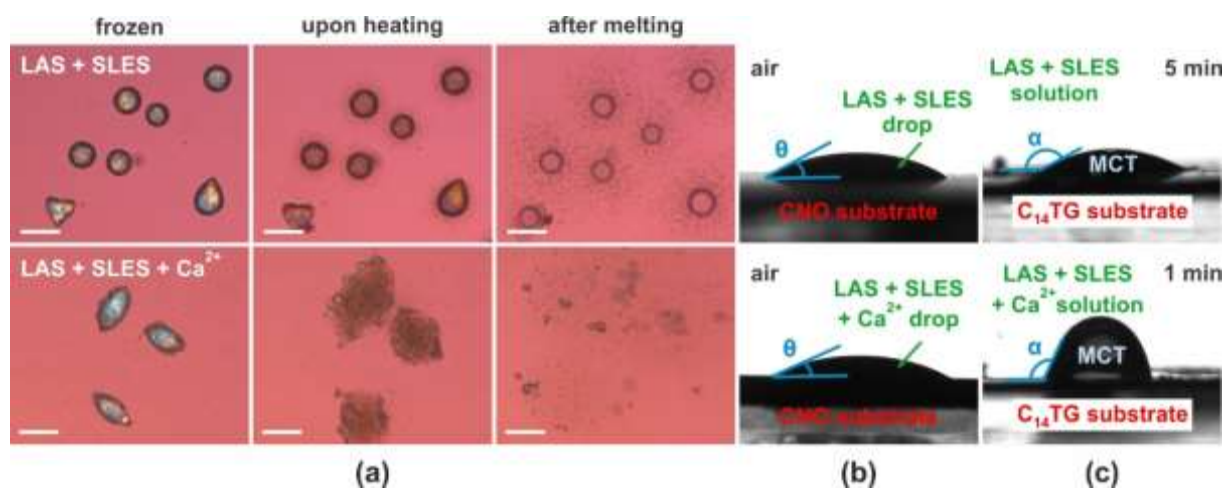

**Figure 8.** **Cold-bursting of CNO particles in LAS + SLES ± $Ca^{2+}$ solutions (3:1 surfactant ratio, 1.5 wt. % total concentration, 30 mM $Ca^{2+}$).** (a) Microscopy images illustrating the cold-bursting observed with CNO particles in the absence (top row) and in the presence (bottom row) of $Ca^{2+}$. The volume averaged drop size after one cooling-heating cycle remains almost unchanged in the sample without $Ca^{2+}$, although numerous small droplets have dewetted from the particle periphery (score 1.5). Complete disintegration is observed in the presence of $Ca^{2+}$ (score 5). Scale bars, 50 μm. (b) Images of drops of the surfactant solutions, placed on a frozen CNO substrate. The angle at the water-air-CNO contact line is $\theta \approx 33°$ for the sample without $Ca^{2+}$ and $\theta \approx 27°$ for the sample with $Ca^{2+}$ at 10°C. (c) Images of MCT drops placed on frozen $C_{14}TG$ substrates in the two surfactant solutions. The angle measured through the aqueous phase with LAS + SLES remains very high even after 5 min of immersion, $\alpha \approx 155°$, while it becomes $\alpha \approx 110°$ after 1 min in the solution with $Ca^{2+}$.

In agreement with these experimental results and the proposed cold-bursting mechanism, the solutions in the absence of $Ca^{2+}$ were completely clear at all temperatures studied, whereas those with calcium were opalescent at low temperatures (they become clear only at $T > 15°C$).

To understand better why LAS + SLES + $Ca^{2+}$ system performs so well, we also measured the other relevant three-phase contact angle, α, i.e. the angle at the contact line "frozen CNO-just melted CNO-surfactant solution", see Figure 3d. Because it is not possible to prepare solid CNO substrate and to place a liquid CNO drop on top of it at a given temperature, we used as model substrate trimyristin ($C_{14}TG$) which resembles the high-melting-temperature components in the CNO. As a liquid oily drop we used medium-chain triglycerides (MCT) which model the low-melting-temperature fraction in CNO. After placing the MCT oil drop on top of the $C_{14}TG$ substrate, we pour the surfactant solution so that the substrate and the MCT drop became entirely immersed in this solution. Thus we measured the three-phase contact angle formed at the $C_{14}TG/MCT/(LAS + SLES \pm Ca^{2+})$ contact line, Figure 3d.

The results from these experiments are shown in Figure 8c. For the system without $Ca^{2+}$, the three phase contact angle, $α$, became *ca.* 155° after the surfactant solution is poured and remains around this value for a period of ≈ 5 min. In contrast, in the system with $Ca^{2+}$, a significant dewetting of the substrate by the oil drop was observed and the contact angle became ≈ 110° within 1 min.

Therefore, the low contact angle θ at the CNO-solution-air contact line, in combination with low contact angle α at the frozen oil-melted oil-water line, and low $σ_{aw}$ with fast surfactant adsorption, characterized by rapidly decreasing σ(*t*), explain the very efficient cold-bursting in the LAS + SLES + $Ca^{2+}$ solution. We note that in this case no oil-soluble surfactant is needed, because the LAS+SLES micelles in the presence of calcium are large non-spherical molecular aggregates.

Finally, we tested SLES + CAPB surfactant combination at 2:1 weight ratio and 1 wt. % total surfactant concentration, with various cosurfactants added at ten times lower concentration of 0.1 wt. %. Previous studies showed that, depending on the added cosurfactant, the shape and size of the SLES+CAPB micelles change significantly.[49] Based on the results obtained with the other surfactants, we hypothesized that such variation in the micelle shape and size could cause significant osmotic effects and could intensify the cold-burst process. Tested cosurfactants included fatty acids with chain lengths varied between $C_8$ and $C_{14}$, and fatty alcohols with chain lengths varied between $C_8$ and $C_{12}$. The results are summarized in Supplementary Figure S9.

With the system SLES + CAPB without any cosurfactant, some separation of small droplets from the periphery of the frozen CNO particles was observed, but the volume averaged drop size after one cooling-heating cycle remained practically unchanged. Significant differences, however, were observed in the presence of additives. As seen from the images in Supplementary Figure S9, all of the solutions with additives with $C_n > C_8$ appeared turbid and, as expected, we observed significant self-dispersion process with all of them, see the scores presented in Supplementary Figure S9.

From all obtained results with CNO particles we can conclude that the following requirements should be fulfilled to observe efficient the cold-burst process:

(1) The three phase contact angle at the air-water-oil interface should be low (*ca.* ≤ 30°). Such low value of the contact angle is most probably needed for rapid penetration of the aqueous phase into the branched channel network formed in the frozen lipid particles.[50,51]

(2) The equilibrium surface tension should be low (*ca.* ≤ 30 mN/m) and the dynamic surface tension $\sigma(t)$ should decrease rapidly. The latter kinetic effect could be overcome in part using much lower heating rates to ensure sufficient time for surfactant adsorption.

(3) If the surfactant solution contains non-spherical micelles (ellipsoidal or bigger supramolecular aggregates) and the preferential adsorption of one of the surfactants can transform these micelles into spherical ones, strong osmotic effects are induced and the cold-burst process becomes very efficient.

(4) Low three phase contact angle measured at the contact line (frozen oil-melted oil-water) also boosts the cold-burst efficiency, facilitating the complete separation of the low- and high-melting fractions of the oil.

The above requirements could be fulfilled using one or a combination of several surfactants (water-soluble and/or oil-soluble) and could be affected by the presence of electrolytes.

### 3. Cold-bursting with other mixed di-/triglyceride oils

To test whether the conclusions reached in the experiments with CNO are valid for other mixed oils (di- and triglycerides), we performed several experiments with palm kernel oil (PKO), cocoa butter (CB), Gelucire 43/01 (GEL01) and Precirol ATO 5 (PRE). The alkyl chain length composition of these oils is presented in Supplementary Table S1.

PKO has similar triglyceride composition to coconut oil. The main difference between PKO and CNO is that PKO has slightly lower content of $C_8$ and $C_{10}$ chains at the expense of a higher fraction of $C_{18:1}$ chain (*ca.* 15 % in PKO vs 6 % in CNO). The peak melting temperature of PKO is $T_m \approx 25°C$, but the complete melting is observed at *ca.* 30°C.

CB contains predominantly $C_{16}$, $C_{18}$ and $C_{18:1}$ alkyl chains and has peak melting temperature ≈ 20°C, with complete melting observed at ≈ 29°C. GEL01 has similar composition to CNO but with slightly longer alkyl chains and without the shortest chains ($C_6$ and $C_8$). The main melting peak is at $T_m \approx 40°C$, with a shoulder observed up to 46.5°C. PRE is a mixture of mono-, di- and triglycerides (MG:DG:TG = 21:54:25) with $C_{16}$ and $C_{18}$ chains, the melting is observed at $T_{m,peak}$ ≈ 48.5°C and ≈ 56°C (double peak).

CNO, PKO and CB are used in food and confectionary industries, while PRE and GEL01 are used in pharmacy and cosmetics. Therefore, demonstrating that the cold-burst phenomenon is applicable to these oils would expand significantly the area of the process application.

Illustrative results with drops of these oils, dispersed in $C_{12}SorbEO_{20}$ + $C_{18:1}MG$ surfactant solution, are presented in Figure 9. As seen from these microscopy images, complete disintegration is observed for PKO (Figure 9a), GEL01 (Figure 9c) and PRE (Figure 9d) oils, whereas with CB (Figure 9b) we do observe some cold-bursting but most of the drops remain with micrometer size.

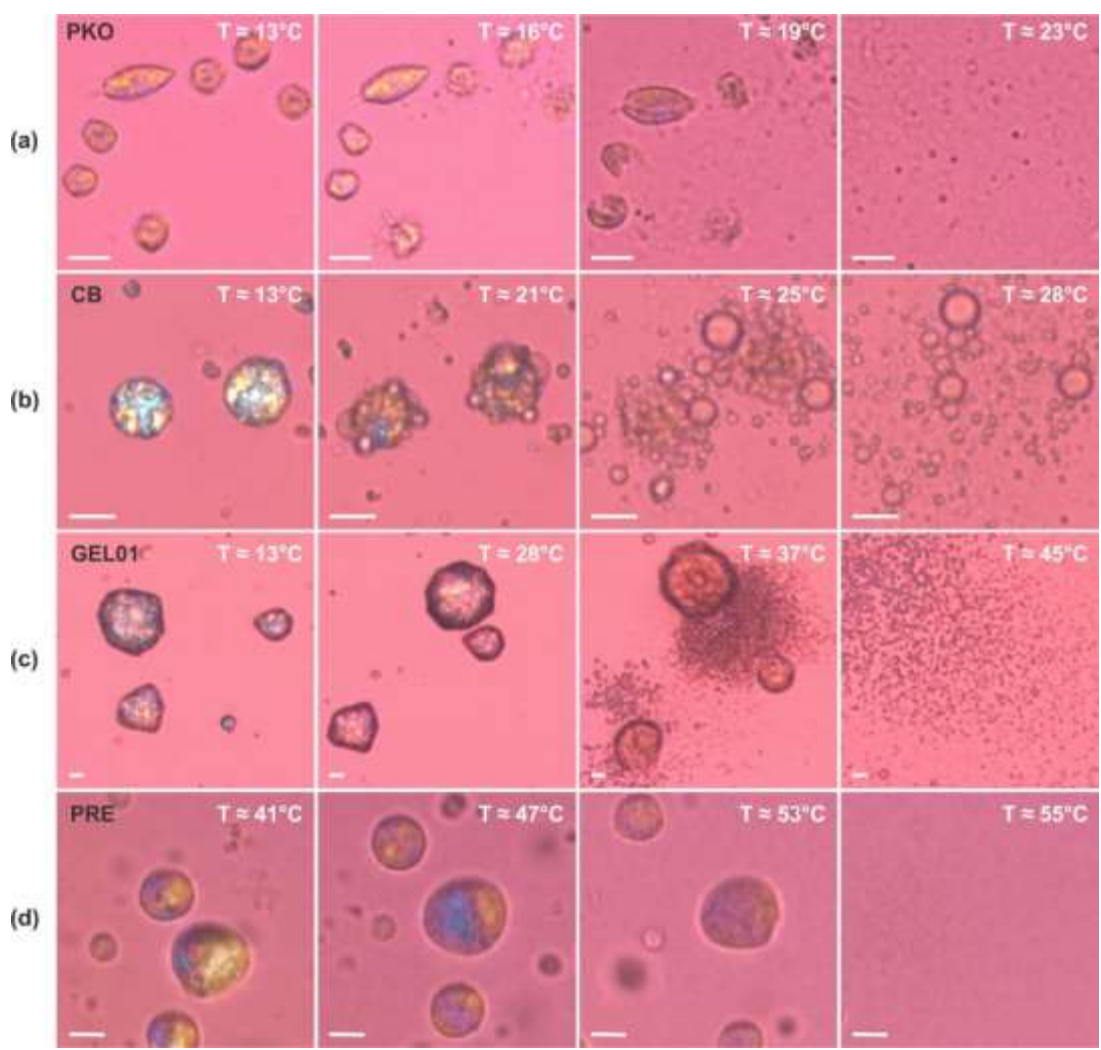

**Figure 9. Cold-bursting process observed with various mixed glyceride oil droplets dispersed in 1.5 wt. % $C_{12}SorbEO_{20}$ and 0.5 wt. % $C_{18:1}MG$.** (a) Palm kernel oil. (b) Cocoa butter. (c) Gelucire 53/01. (d) Precirol ATO 5. Drop with initial size ≈ 10 μm from PKO, PRE and GEL01 disintegrate completely into numerous submicrometer droplets. The disintegration process is also observed for CB is not so efficient under this temperature protocol (cooling to 0°C and heating until complete melting). Scale bars, 10 μm.

The most efficient drop disintegration was observed with PRE, probably because this oil contains saturated $C_{16}$ and $C_{18}$ alkyl chains only. In contrast, CB oil contains a high fraction of unsaturated alkyl chains (ca. 34 %). For that reason, in the DSC thermogram, the melting peak of CB starts at *ca.* 0°C and finishes at 29°C, while the freezing peak starts at *ca.* 20°C and continues down to -18°C. The heating presented in Figure 9b started from an initial temperature of ≈ 3°C. Therefore, we hypothesized that the less efficient disintegration of CB may be due to the fact that a fraction of this oil had not been completely frozen in the emulsions cooled to ca. 0°C. To test this hypothesis, we prepared CB sample dispersed in $C_{12}SorbEO_{20}$ + $C_{18:1}MG$ solution in presence of 50 wt. % ethylene glycol. Then we stored this sample in a glass bottle at -18°C overnight and made the heating the microscope microscopy observation in the next day. The disintegration observed in this case was much more pronounced, similar to the one with GEL01.

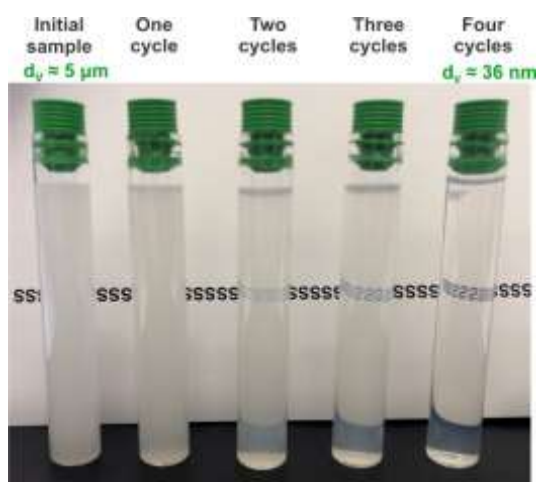

**Figure 10. Cold-bursting of bulk samples with 1 wt. % PRE drops dispersed in 1.5 wt. % $C_{18}EO_{20}$ + 0.5 wt. % $C_{12}EO_4$ solution.** Initial sample contains particles with diameter ≈ 5 μm and appears milky white. The drop size decreases down to 250 nm after one cooling-heating cycle. The main part of the drops disintegrate into the smallest possible size, *ca.* 20 nm after 2 consecutive cycles, however some small fraction of particles remain with higher sizes making the emulsions opalescent. After four consecutive cycles > 99 % of the drops achieve size $d_V$ ≈ 20 nm and the emulsion become completely transparent (some small fraction of the particles remained with bigger sizes, increasing the averaged size measured by volume to 36 nm).

Performing experiments with bulk samples, we found that the smallest drop size with PRE oil was obtained with PRE (1 wt. % emulsion), starting from drops with an initial size of ≈ 5 μm, dispersed in $C_{18}EO_{20}$ + $C_{12}EO_4$ solution. With this emulsion, we obtained $d_V$ ≈ 250 ± 50 nm and $d_N$ ≈ 15 ± 5 nm after one cooling-heating cycle, while $d_V$ ≈ 36 ± 15 nm and $d_N$ ≈ 15 nm were measured with the same sample after 4 consecutive cooling-heating cycles. These samples were completely transparent, Figure 10. We expect that, after further optimization of the temperature protocols, such small sizes could be achieved using smaller number of cycles.

**CONCLUSIONS**

The current study demonstrates that the cold-burst phenomenon observed previously with pure triglycerides and with simple binary or ternary mixtures of monoacid triglycerides is applicable also to much more complex triglyceride mixtures with various alkyl chains, incl. coconut oil, palm kernel oil, cocoa butter, Precirol ATO 5 and Gelicure 43/01.

The spontaneous bursting upon heating of frozen oily particles, dispersed in aqueous surfactant solution, is particularly efficient when the following requirements are fulfilled:

(1) The surfactants ensure that three-phase contact angle at the air-water-oil contact line is low ≤ *ca.* 30°;

(2) The equilibrium surface tension of the surfactant solution is low ≤ *ca.* 30 mN/m and the rate of adsorption is relatively fast;

(3) The three phase contact angle at the melted oil-frozen oil-water interface is below *ca.* 150°; and/or

(4) The aqueous solution contains large micellar aggregates which transform into spherical micelles when some of the surfactant components adsorbs on the surface of the nanopores in the lipid particles, thus inducing strong osmotic effect which sucks water into the particle interior.

The most efficient cold-burst protocol is rapid cooling, followed by slow heating. In this protocol the high cooling rates ensure the formation of smaller crystallites in the frozen lipid particles, whereas the slower heating ensures longer time for penetration of the aqueous phase into the porous network, surfactant adsorption and dewetting of the liquid oil fraction from the still frozen one.

For the mixtures of water- and oil-soluble surfactants, the cold-burst process is observed only if there is a sufficiently high fraction of oil-soluble surfactant dispersed in the aqueous phase which ensures the formation of large micellar aggregates and the osmotic effect described above.

With the obtained new results and their mechanistic explanation we have opened the door for application of the cold-burst method in numerous applied areas. On the basis of the guiding principles, formulated in this study, one could approach in a rational way any specific system of interest – to optimize the surfactants and the cooling-heating protocol in a way ensuring the formation of nanoparticles with a desired size range, using a minimum number of cooling-heating cycles. These guiding principles are expected to apply to wide range of surfactant-oil combinations, because they are based on rigorous physicochemical understanding of the phenomena involved and include the respective key physicochemical parameters – the contact angles θ and α, equilibrium and dynamic surface tensions, aggregation number of the surfactant aggregates and the corresponding osmotic pressure of the surfactant solutions.

**Acknowledgements**

The authors thank Ms. Anita Biserova, Sofia University, for her help with part of the experimental work. The commercial grade monoolein fractionation and GC analysis were performed by Mr. Delyan Krastev, Mrs. Mariana Boneva-Astrukova and Dr. Zahari Vinarov Sofia University (using a procedure defined by S.T. and D.C) – their help is gratefully acknowledged. The authors thank Mrs. Dora Dimitrova for measuring the surface tensions. This study was partially funded by the Proof-of-Concept grant CoolNanoDrop (#841827). The authors gratefully acknowledge the support from the Operational Program "Science and Education for Smart Growth", Bulgaria, grant number BG05M2OP001-1.002-0023.

**Author contributions:**

S.T., N.D. and D.C. designed the study; D.G. performed most of the experiments, summarized the results, and prepared part of the figures and movies; D.C. and S.T. analyzed the results; D.C., S.T. and N.D. clarified the mechanisms; D.C. prepared the first draft of the manuscript, whereas N.D. edited it and prepared the final draft; S.T. read critically the manuscript and suggested improvements; All authors participated in discussions and critically read the final manuscript.

**Supplementary Information.** Movies showing the cold-bursting process with $C_{12}TG$ (Movie 1) and CNO (Movies 2 and 3); Alkyl chain length composition of the studied natural triglyceride oils (Table S1); Surfactant structures and properties (Table S2); Equilibrium surface tensions and three phase contact angles for various systems (Table S3); Visual appearance of mixed surfactant solutions (Figure S1 and S9); Temperature dependence of the contact angles (Figure S2); Microscopy images for bulk samples (Figure S3); Microscopy images for cold-bursting with different systems (Figures S4 and S6); Dynamic surface tensions (Figure S5 and S7); Effect of the heating rate (Figure S8);